\documentclass[aps,prl,reprint,superscriptaddress,amsmath,amssymb,showpacs,floatfix]{revtex4-1}
\usepackage{graphicx}
\usepackage{bm}
\usepackage{color}
\usepackage[applemac]{inputenc}
\usepackage[T1]{fontenc}

\begin{document}

\title{Three heats in strongly coupled system and bath}

\author{Chulan Kwon}
\affiliation{Department of Physics, Myongji University, Yongin, Gyeonggi-Do,
17058,  Korea}
\email{ckwon@mju.ac.kr}
\author{Jaegon Um}
\affiliation{BK21PLUS Physics Division, Pohang University of Science and Technology, Pohang 37673, Korea}
\email{slung@postech.ac.kr}
\author{Joonhyun Yeo}
\affiliation{Department of Physics, Konkuk University, Seoul 05029, Korea}
\author{Hyunggyu Park}
\affiliation{School of Physics, Korea Institute for Advanced Study, Seoul 02455, Korea}

\date{\today}

\begin{abstract}
We investigate three kinds of heat produced in a system and a bath strongly coupled via an interaction Hamiltonian. By studying the energy flows between the system, the bath, and their interaction, we provide rigorous definitions of two types of heat, $Q_{\rm S}$ and $Q_{\rm B}$ from the energy loss of the system and the energy gain of the bath, respectively. This is in contrast to the equivalence of  $Q_{\rm S}$ and $Q_{\rm B}$, which is commonly assumed to hold in the weak coupling regime. The bath we consider is equipped with a thermostat which enables it to reach an equilibrium. We identify another kind of heat $Q_{\rm SB}$ from the energy dissipation of the bath into the super bath that provides the thermostat. We derive the fluctuation theorems (FT's) with the system variables and various heats, which are discussed in comparison with the FT for the total entropy production. We take an example of a sliding harmonic potential of a single Brownian  particle in a fluid and calculate the three heats in a simplified model. These heats are found to equal on average in the steady state of energy, but show different fluctuations at all times.
\end{abstract}
\pacs{05.70.Ln, 02.50.-r, 05.40.-a}
\maketitle
The nonequilibrium fluctuation theorem (FT) has been proven originally for deterministic
systems~\cite{evans93,evans94,gallavotti}, later for stochastic
systems~\cite{jarzynski1,jarzynski2,crooks,kurchan,lebowitz,seifert,esposito},
and recently for quantum systems~\cite{campisi,campisi_RMP,hanggi,talkner}. It
takes into account thermodynamic quantities such as heat and work which are continuously produced
even in the steady state. Such quantities accumulated for a
long time exhibit huge fluctuations around their means, which is especially prominent in small
systems. Compared to work, heat is intriguing because it is interpreted as an energy exchange with practically unrecognizable bath. By assuming the master equation or the Langevin equation, heat is found as a function of stochastic trajectories~\cite{schnakenberg,sekimoto}.

Recent studies, mostly quantum mechanical, have more concentrated on a system strongly coupled with a bath~\cite{nieuwenhuizen,hoerhammer,ilki_kim,esposito_qm1,pucci,horowitz1,hekking,horowitz2,silaev,
gallego,ankerhold,carrega,esposito_qm2,seifert_strong_coupling,esposito_strong,iyoda_sagawa,funo1}.
In spite of extensive efforts, however, it is pointed out in Ref.~\cite{hanggi} that a consistent
definition of heat for the strong coupling regime is currently not known and most of the
studies are restricted to the assumption of an initial product state.
A proper means to treat the interaction energy or Hamiltonian between the system and the bath
is still missing. This limitation is also present in classical approaches.
In this study, we develop a theoretical framework to deal with the interaction
rigorously for strongly coupled classical systems, which is expected to extend to quantum systems.

As the interaction energy changes in time, it accompanies energy changes
in both system and bath, hence we expect two different forms of heat. We consider the bath to be equipped with a thermostat provided by another external system, which we call a super bath, so that the total system and bath is able to reach an equilibrium in the absence of a nonequilibrium source.
We then find another form of heat in the bath, which is dissipated into the super bath and plays a crucial
role to prevent the bath from heating up indefinitely.

In this paper, we present detailed mathematical definitions for the three heats
based on the rigorous treatment of the interaction Hamiltonian.
We show that there are many different versions of the FT for entropy production
due to the three forms of heat. We find that three heats exhibit different fluctuations even in the steady state. We explicitly confirm these properties from a specific example.

First, we consider a general particle Hamiltonian system and bath coupled.
The system variables are given by a collection of momentums ${\vec p}=({\vec p}_1, {\vec p}_2, \ldots,)^{\rm t}$
and positions ${\vec x}=({\vec x}_1, {\vec x}_2, \ldots)^{\rm t}$, where the superscript ${\rm t}$ denotes the transposition of vector or matrix. Similarly, the bath variables are given
by ${\vec p}_{\rm B}$ and ${\vec x}_{\rm B}$.
The Hamiltonian of the total system is composed of three parts:
$H_{\rm S}={\vec p}^2/(2\mu)+U({\vec x},\lambda(t))$ for the system,
$H_{\rm B}={\vec p}_{\rm B}^2/(2m)+U_{\rm B}({\vec x}_{\rm B})$ for the bath, and
$H_{\rm I}= V({\vec x}, {\vec x}_{\rm B})$ for the interaction,
where the time-dependent protocol $\lambda(t)$
is prescribed only in the system potential $U$  and the interaction Hamiltonian $V$  is a pairwise potential between the system and bath particles. We take the same mass $\mu$ for all system particles and $m$ for all bath particles, just for notational convenience.
We assume that the bath is equipped with a Langevin thermostat provided by super bath.
Then, equations of motion read as
$\dot{\vec x}=\partial H_{\rm S}/\partial {\vec p}$, $\dot{\vec p}=-\partial(H_{\rm S}+ H_{\rm I})/\partial {\vec x}$,
$\dot{\vec x}_{\rm B}=\partial H_{\rm B}/\partial {\vec p}_{\rm B}$, and $\dot{\vec p}_{\rm B}=-\partial(H_{\rm B}+ H_{\rm I})/\partial {\vec x}_{\rm B}-\gamma {\vec p}_{\rm B}/m+{\vec \xi}(t)~$ with
the white noise ${\vec \xi}(t)$ satisfying $\langle \xi_i(t)\xi_j(t')\rangle=2\gamma\beta^{-1}\delta_{ij}\delta(t-t')$
for the inverse temperature $\beta$ and the viscosity coefficient $\gamma$.

Equations of motion lead to energy relations:
\begin{equation}
\frac{{\rm d}H_{\rm S}}{{\rm d}t}=\dot{W}-\dot{Q}_{\rm S},
~ \frac{{\rm d}(H_{\rm S}+H_{\rm I})}{{\rm d}t}=\dot{W}-\dot{Q}_{\rm B},
~\frac{{\rm d}H}{{\rm d}t}=\dot{W}-\dot{Q}_{\rm SB}~.
\label{H_tot}
\end{equation}
where $\dot{W}=\partial U/\partial t$ is the rate of the work produced by the time-dependent protocol $\lambda(t)$,  $\dot{Q}_{\rm S}$ ($\dot{Q}_{\rm B}$) the rate of heat loss (gain) of the system (bath).  $\dot{Q}_{\rm SB}$ is the rate of heat loss of the bath flowing into the super bath (SB) surrounding it, which was also considered in a recent study~\cite{esposito_strong}.  The rates of the three heats are defined as
\begin{equation}
\dot{Q}_{\rm S}=\frac{\partial V}{\partial {\vec x}}\cdot \frac{\vec p}{\mu},~
\dot{Q}_{\rm B}=-\frac{\partial V}{\partial {\vec x}_{\rm B}}\cdot \frac{{\vec p}_{\rm B}}{m},~
\dot{Q}_{\rm SB}=\left( \frac{\gamma  {\vec p}_{\rm B}}{m}-{\vec \xi}\right)\cdot \frac{{\vec p}_{\rm B}}{m}.\label{3heats}
\end{equation}
Note that ${\rm d}H_{\rm I}/{\rm dt}=\dot{Q}_{\rm S}- \dot{Q}_{\rm B}\neq 0$, which is contrary to the usual expectation about the heat exchange between system and bath. From the bath point of view, we get ${\rm d}H_{\rm B}/{\rm dt}=\dot{Q}_{\rm B}-\dot{Q}_{\rm SB}$,
where the thermostat slows down the increase of the bath energy.
For equilibrium bath, the driving on the system
by the time-dependent protocol should be mild enough to maintain the bath energy saturated in the long-time limit. Even in this case, the $\dot{Q}_{\rm B}$ and $\dot{Q}_{\rm SB}$ may show different fluctuations.

Now, we examine the FT for our model. Though the Langevin thermostat is connected partially only to the bath, the total system plus bath are governed by the Langevin dynamics for which various forms of FT are already known to hold~\cite{kurchan,lebowitz}. For example,
the integral FT  holds for the total entropy production $\Delta S$ accumulated during a finite time interval as
\begin{equation}
\langle e^{-\Delta S}\rangle =1 \quad {\rm with} \quad \Delta S=-\Delta \ln \rho+\beta Q_{\rm SB}~,
\label{TFT}
\end{equation}
where $-\Delta \ln \rho$ represents the Shannon entropy change with $\rho$
the probability distribution function (PDF) of the total system and
$Q_{\rm SB}$ the accumulated heat flowing into the super bath. Here and throughout our paper, we set the Boltzmann constant $k_{\rm B}=1$.

These trivial FT's  are, however, not very informative from the system point of view. Let
$\mathbf{q}=({\vec x}, {\vec p}, {\vec x}_{\rm B}, {\vec p}_{\rm B})^{\rm t}$ be a
state vector of the total system with $\mathbf{q}_{\rm S}=({\vec x}, {\vec p})^{\rm t}$ for system
and $\mathbf{q}_{\rm B}=({\vec x}_{\rm B}, {\vec p}_{\rm B})^{\rm t}$ for bath, respectively.
The reduced system PDF $\rho_{\rm S}(\mathbf{q}_{\rm S})$ defined as ${\rm Tr}_{\rm B} \rho(\mathbf{q})$
is obtained by tracing out the bath variable $\mathbf{q}_{\rm B}$
for the total system PDF $\rho(\mathbf{q})$. Then, the Bayes' rule
leads to $\rho(\mathbf{q})=\rho_{\rm S}(\mathbf{q}_{\rm S}) {\rho}(\mathbf{q}_{\rm B}|\mathbf{q}_{\rm S})$ with ${\rho}(\mathbf{q}_{\rm B}|\mathbf{q}_{\rm S})$ the conditional PDF.

In deriving the FT for the system variables, it is useful to
introduce a reference state for the total system. In this paper, we consider two typical reference states in the form of
$\tilde{\rho}(\mathbf{q})=\rho_{\rm S}(\mathbf{q}_{\rm S})\tilde{\rho}(\mathbf{q}_{\rm B}|\mathbf{q}_{\rm S})$, where the
conditional PDF's for the two cases are given by
\begin{eqnarray}
\textrm{(a)}&&~~\tilde{\rho}(\mathbf{q}_{\rm B}|\mathbf{q}_{\rm S})= Z_{\rm B}^{-1}e^{\beta\tilde{H}_{\rm S}}e^{-\beta (H_{\rm B}+H_{\rm I})}, \nonumber\\
\textrm{(b)}&&~~\tilde{\rho}(\mathbf{q}_{\rm B}|\mathbf{q}_{\rm S})=Z_{\rm B}^{-1}e^{-\beta H_{\rm B}},
\label{Ref}
\end{eqnarray}
where the equilibrium bath partition function $Z_{\rm B}={\rm Tr}_{\rm B} e^{-\beta H_{\rm B}}$ and the {\em additional} system Hamiltonian $\tilde{H}_{\rm S}(\mathbf{q}_{\rm S})$
originates from the normalization as $e^{-\beta \tilde{H}_{\rm S}}=Z_{\rm B}^{-1} {\rm Tr}_{\rm B} e^{-\beta (H_{\rm B}+H_{\rm I})}$. One can see easily that
$\tilde{H}_{\rm S}$ vanishes in the limit  
$H_{\rm I} \approx 0$.

The case
(a) is a special type recently considered by Seifert~\cite{seifert_strong_coupling}.
If the total system is in equilibrium, (a) is exact and the reduced system PDF becomes
$\rho_{\rm S} (\mathbf{q}_{\rm S})\sim e^{-\beta H_{\rm S}^{\rm eff}}$
with $H_{\rm S}^{\rm eff}=H_{\rm S} + \tilde{H}_{\rm S}$, indicating that the strong coupling
induces an additional term in the system Hamiltonian.
The case (b) corresponds to the usual assumption of the product state of system and bath.

Difference between the true and reference states can be measured by the
relative entropy as $D(\rho||\tilde{\rho})=\ln[\rho/\tilde{\rho}]$.
Then, the total entropy production $\Delta S$ in Eq.~\eqref{TFT} can be rewritten
in terms of the system PDF and other energy variables along with the relative entropy
such as
\begin{eqnarray}
\textrm{(a)} &&~\Delta S = -\Delta \ln\rho_{\rm S}+\beta (Q_{\rm S}-\Delta \tilde{H}_{\rm S})-\Delta D_{\rm a}, \nonumber\\
\textrm{(b)} &&~\Delta S =-\Delta \ln\rho_{\rm S} +\beta Q_{\rm B}-\Delta D_{\rm b},
\label{TFT1}
\end{eqnarray}
where $\Delta D_{\rm a, b}$ are the relative entropy changes for the types (a) and (b), respectively.
In this derivation, we utilized energy relations as $Q_{\rm SB}-Q_{\rm B}=-\Delta H_{\rm B}$ and
$Q_{\rm SB}-Q_{\rm S}=-\Delta (H_{\rm B}+H_{\rm I})$.
Then, the thermodynamic second laws yield the inequalities,
\begin{eqnarray}
\textrm{(a)}&&~R_a=\langle-\Delta \ln\rho_{\rm S}+\beta (Q_{\rm S}-\Delta \tilde{H}_{\rm S})-\Delta D_{\rm a}\rangle \ge 0 ,\label{ineq_a}\nonumber\\
\textrm{(b)}&&~R_b=\langle -\Delta \ln\rho_{\rm S}+\beta Q_{\rm B}-\Delta D_{\rm b}\rangle \ge 0 ,\label{ineq1}
\end{eqnarray}
where the equality holds for non-thermostatted bath ($Q_{\rm SB}=0$), due to $\Delta \ln\rho=0$ for the Louiville dynamics. Nevertheless, the FT's and second laws in the above forms still
require the knowledge of the relative entropy change which cannot be accessible without knowing the true PDF ${\rho}(\mathbf{q}_{\rm B}|\mathbf{q}_{\rm S})$ of the bath. 

One can get around this when the initial state is not arbitrary but of our reference state
(a) or (b) in Eq.~\eqref{Ref}. For example, consider a quantity
$\Delta A\equiv -\Delta \ln\rho_{\rm S}+\beta (Q_{\rm S}-\Delta \tilde{H}_{\rm S})$, appeared in Eq.~\eqref{TFT1}, which does not require the knowledge of the bath PDF. With the initial
condition prepared with the reference state (a), we get, for a finite time interval $t=[0,\tau]$,
\begin{eqnarray}
&&\langle e^{-\Delta A} \rangle_{\rm a}
=\int\!\!{\rm D}\mathbf{q}(t)~
e^{-\Delta A}  \Pi[\mathbf{q}(t);\lambda(t)]
\frac{\rho_{\rm S}(0) e^{-\beta (H_{\rm B}(0)+H_{\rm I}(0))}}{Z_{\rm B}e^{-\beta\tilde{H}_{\rm S}(0)}}\nonumber\\
&&=\int\!\!{\rm D}\mathbf{q}_{\rm R}(t)~
 \Pi[\mathbf{q}_{\rm R}(t);\lambda_{\rm R}(t)] \frac{ \rho_{\rm S}(\tau) 
e^{-\beta (H_{\rm B}(\tau)+H_{\rm I}(\tau))}}{Z_{\rm B}e^{-\beta\tilde{H}_{\rm S}(\tau)}} = 1, 
 \label{FTA}
\end{eqnarray}
where $ \Pi[\mathbf{q}(t);\lambda(t)]$ ($ \Pi[\mathbf{q}_{\rm R}(t);\lambda_{\rm R}(t)]$) is the standard conditional probability for the path (reverse path) $\mathbf{q}(t)$ ($\mathbf{q}_R(t))$ and the protocol (time-reversed protocol) is $\lambda(t)$ ($\lambda^R(t)$), and the Schnakenberg relation $\Pi[\mathbf{q}(t);\lambda(t)]/\Pi[\mathbf{q}_R(t);\lambda^R(t)]=e^{\beta Q_{\rm SB}}$ is used.
Note that $\langle\cdots\rangle_{\rm a}$ is the average with the initial state of reference type (a). The final equality comes from the probability normalization because the second integral represents
the sum of all possible paths in the reverse process with its initial state of the same
reference type (a).
Similarly, we get for $\Delta B\equiv -\Delta \ln\rho_{\rm S}+\beta Q_{\rm B}$ as
$\langle e^{-\Delta B}\rangle_{\rm b} =1$, when the initial condition is prepared with the reference state (b). The FT for $\Delta A$  in Eq.~\eqref{FTA} has been recently found by Seifert~\cite{seifert_strong_coupling} in the case without the super bath, and the FT for $\Delta B$
has been known for quantum systems~\cite{iyoda_sagawa}. 

\begin{figure}
\centering
\includegraphics*[width=\columnwidth]{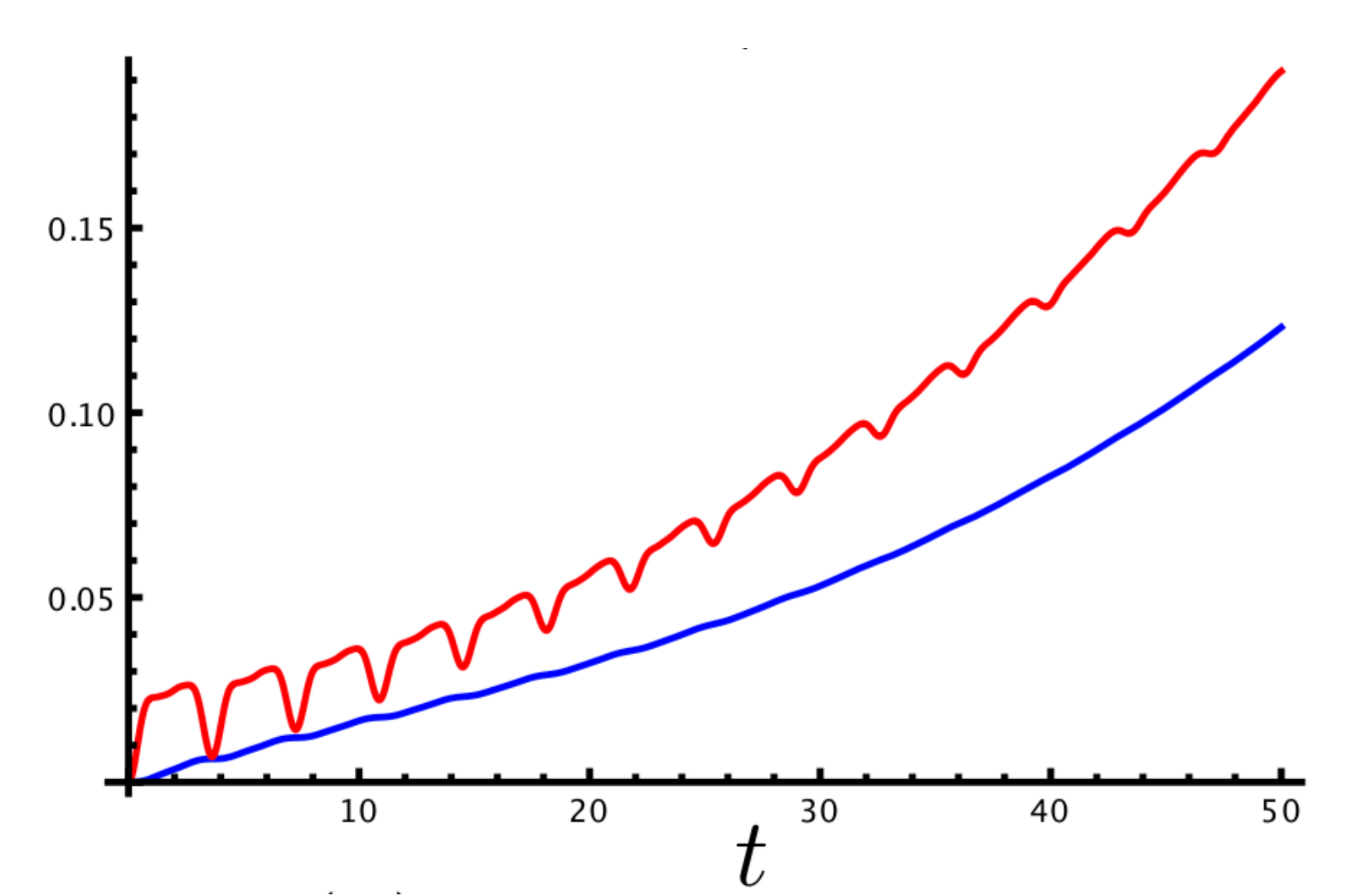}
\caption{(Color online) $R_A$ (red, upper) and $R_a$ (blue, lower) versus time $t$ for $\beta'=1.5$, $u=0.02$. We use $N=2$, $\gamma=30$, $\mu=m=1$, $k=\kappa=1$, and $\beta=1$. $R_A>R_a$ and $R_A$ is not monotonous for this weak nonequilibrium case, as expected.}
\label{fig1}
\end{figure}

The corresponding inequalities are given as
\begin{eqnarray}
\textrm{(A)}&&~R_A=\langle-\Delta \ln\rho_{\rm S}+\beta (Q_{\rm S}-\Delta \tilde{H}_{\rm S})\rangle_{\rm a} \ge 0 ,\label{ineq_a}\nonumber\\
\textrm{(B)}&&~R_B=\langle -\Delta \ln\rho_{\rm S}+\beta Q_{\rm B}\rangle_{\rm b} \ge 0 .
\label{ineq2}
\end{eqnarray}
One should notice that $R_A$ or $R_{B}$ do not necessarily increase with interval time $\tau$
($dR_A/d\tau$ and $dR_B/d\tau$ can be negative), because the total system PDF does not maintain its form of reference states as soon as the evolution starts. In contrast, $R_a$ or $R_b$ should increase always with $\tau$. These properties will be shown explicitly from rigorous calculations for a simple example later shown in Fig.~\ref{fig1}.
We remark that, with reference initial states, $R_A \ge \langle D_a(\tau)\rangle_a
\ge 0$ from Eq.~\eqref{ineq1}, which implies the inequality for the total entropy production 
provides a tighter bound by the amount of the relative entropy at the final time. A similar result was found in Ref.~\cite{esposito_strong}.

Now, we take a concrete example for explicit calculation of three heats in average and also their PDF's.
Consider a Brownian colloidal particle submerged in a fluid bath. This colloid interacts with bath particles nearby through a finite-range interaction. These perturbed bath particles relax fast into equilibrium and new bath particles begin to interact as the colloid moves through the bath.
For an analytic approach, we mimic this situation by considering only a small number $N$ of bath particles moving along with the colloid through strong harmonic interactions~\cite{exp1}. All other non-interacting bath particles are in equilibrium.  

For simplicity, we only consider the one-dimensional model and take the bath potential $U_{\rm B}=0$.
The total system state is given by $\mathbf{q}=(x, p, x_1, p_1, \cdots, x_N, p_N)^{\rm t}$ with the system state $\mathbf{q}_{\rm S}=(x,p)^{\rm t}$ and the bath state
$\mathbf{q}_{\rm B}=(x_1, p_1, \cdots, x_N, p_N)^{\rm t}$. We use a different ordering of the components of the state vectors from the previous one. Note that we dropped state variables of all other bath particles which do not interact with the colloid. The interaction Hamiltonian is written as $H_I=\sum_iV_i$ where the interaction potential between the colloid  and the $i$-th bath particle is chosen as $V_i=\kappa(x-x_i)^2/2$, which is long-ranged enough to keep interacting bath particles near the colloid. In order to study non-equilibrium motion, 
we introduce a sliding 
harmonic potential with a constant velocity $u$ given by $U(x,\lambda(t))=k(x-\lambda(t))^2/2$ with $\lambda(t)=ut$.
This protocol for a Brownian particle has been extensively studied experimentally~\cite{wang,hummer,wang05} and theoretically~\cite{vanzon1,vanzon2,kwangmoo}
for a single-particle Langevin system. 

We define $\mathbf{q}^*=\mathbf{q}-\mathbf{u}t$ for $\mathbf{u}=(u,0,u,0,\ldots)^{\rm t}$. Then, the total Hamiltonian at time $t$ can be expressed as a function of $\mathbf{q}^*$ with no explicit time dependence, given as
\begin{equation}
H(\mathbf{q}^*)=\frac{1}{2}\mathbf{q}^*\cdot\mathsf{A}_{\rm eq}\cdot\mathbf{q}^*=\frac{1}{2}\mathbf{q}^*\cdot(\mathsf{A}_{\rm S}+\mathsf{A}_{\rm I}+\mathsf{A}_{\rm B})\cdot\mathbf{q}^*
\end{equation}
where various matrices $\mathsf{A}$ are obtained from the corresponding Hamiltonians $H$, $H_{\rm S}$, $H_{\rm I}$, and $H_{\rm B}$ which are quadratic in $\mathbf{q}^*$.
We decompose $\mathbf{q}^*$ into a stochastic part $\mathbf{z}$ and a deterministic part $\mathbf{d}$, which are governed by
\begin{eqnarray}
\dot{\mathbf{d}}=-\mathsf{F}\cdot\mathbf{d}-\mathbf{u}~,~~
\dot{\mathbf{z}}=-\mathsf{F}\cdot\mathbf{z}+\boldsymbol{\xi}(t)~.\label{separation}
\end{eqnarray}
Here, $\mathsf{F}$ is a $d\times d$ positive-definite matrix
with $d=2(N+1)$. The white noise $\boldsymbol{\xi}(t)$ acts exclusively on the momenta of bath particles. See Sec.~I in the Supplementary Material (SM) for the explicit forms of matrices~\cite{SM}.

We get $\mathbf{d}(t)=-\mathsf{F}^{-1}(\mathbf{I}-e^{-\mathsf{F}t})\cdot\mathbf{u}$ for the initial condition $\mathbf{d}(0)=\mathbf{0}$. The PDF for $\mathbf{z}$ at time $t$ is given~\cite{kwon-ao-thouless,kwon} as
$\sigma(\mathbf{z},t)=\sqrt{\frac{|\beta\mathsf{A}_t|}{(2\pi)^d}} \exp\left[-\frac{\beta}{2}\mathbf{z}^{\rm t}\cdot\mathsf{A}_t\cdot\mathbf{z}~\right]$
where
$\mathsf{A}_t^{-1}=\mathsf{A}_{\rm eq}^{-1}-\mathsf{U}_{t,0}(\mathsf{A}_{\rm eq} -\mathsf{A}_0^{-1}){\mathsf{U}}_{t,0}^{\rm t}$
for $\mathsf{U}_{t,t'}=e^{-\mathsf{F}(t-t')}$. Then, the PDF for $\mathbf{q}$ at $t$ is given by \begin{equation}
\rho(\mathbf{q},t)=\sqrt{\frac{|\beta\mathsf{A}_t|}{(2\pi)^d}} e^{-\frac{\beta}{2}\left[\mathbf{q}-\mathbf{u}t-\mathbf{d}(t)\right]^{\rm t}\cdot\mathsf{A}_t\cdot\left[\mathbf{q}-\mathbf{u}t-\mathbf{d}(t)\right]}.
\label{pdf}
\end{equation}
The nonequilibrium nature of the system is characterized by a nonzero value of $\langle\mathbf{q}\rangle=\mathbf{u}t+\mathbf{d}(t)$.

We write the three heats and work accumulated for $0<t<\tau$ using Eq.~(\ref{H_tot}) as
\begin{eqnarray}
W&=&\int_0^\tau dt\frac{\partial U(x,t)}{\partial t}=-ku\int_0^\tau\!\!\!\! {\rm d}t[\mathbf{q}^*(t)]_x
\label{work}\\
Q_{\alpha}&=&W-\Delta\left[\frac{1}{2}\mathbf{q}^*\cdot \mathsf{B}_\alpha\cdot\mathbf{q}^*\right]
\label{3heats}
\end{eqnarray}
where the subscript $x$ denotes the first (system position) component of the vector and
$\mathsf{B}_{\alpha}=\mathsf{A}_{\rm S}$, $\mathsf{A}_{\rm S}+\mathsf{A}_{\rm I}$, and $\mathsf{A}_{\rm eq}$, respectively for $\alpha={\rm S},~{\rm B},~{\rm SB}$.

We can choose an initial condition according to type (a) such as $\mathsf{A}_0=(\beta'/\beta)\mathsf{A}_{\rm S}+\mathsf{A}_{\rm B}+\mathsf{A}_{\rm I}$ and find $\rho(\mathbf{q},t)$ from Eq.~(\ref{pdf}). Then, we find $R_{a}$ and $R_{A}$ for Fig.~\ref{fig1}; see Sec.~II in SM~\cite{SM}. The behavior of the two quantities in time is presented in Fig.~\ref{fig1}, which is consistent with the expectation. 

\begin{figure}[t!]
\centering
\includegraphics*[width=\columnwidth]{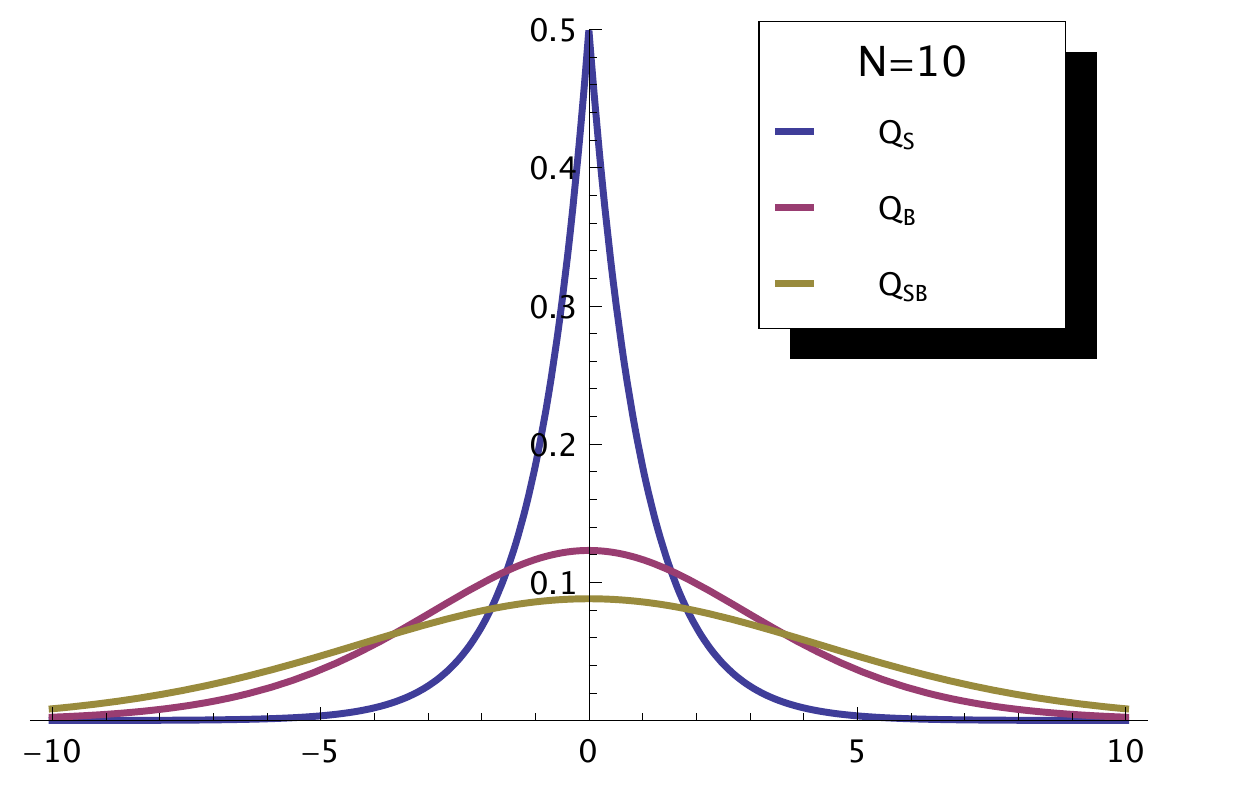}
\caption{(Color online) Plots for ${\cal P}_{\alpha}^{\rm eq}(r)$ for $r=\beta Q$.  The distributions become broader for larger $N$. For the same $N$, ${\cal P}_{\rm SB}^{\rm eq}$ is the broadest and ${\cal P}_{\rm S}^{\rm eq}$ is the sharpest.  ${\cal P}_{\rm S}^{\rm eq}(r)=e^{-|r|/2}$ is independent of $N$.  }
\label{fig2}
\end{figure}

The generating function for the heat distribution is defined as
${\cal G}_\alpha(\lambda)=\langle e^{-\beta \lambda Q_{\alpha}}\rangle$ where $\langle\cdots\rangle$ denotes the average over all trajectories $\mathbf{z}(t)$ for $0<t<\tau$, and the initial and final states, $\mathbf{z}_0$ and $\mathbf{z}_\tau$,
for an initial PDF and the trajectory probability. For convenience, we only consider the $\beta^\prime=\beta$ case for the initial condition, implying that the total system 
is in equilibrium at the beginning. We have
\begin{eqnarray}
{\cal G}_{\alpha}(\lambda)&=&\left\langle e^{\frac{\beta\lambda}{2}{\mathbf{q}_\tau^*}^{\rm t}\cdot\mathsf{B}_{\alpha} \cdot\mathbf{q}_\tau^*}
e^{\beta\lambda ku \int_0^\tau{\rm d}t [\mathbf{q}^*(t)]_x}
e^{-\frac{\beta\lambda}{2}{\mathbf{q}_0^*}^{\rm t}\cdot\mathsf{B}_{\alpha} \cdot\mathbf{q}_0^*}\right\rangle \nonumber\\
&=&c_\alpha N_\alpha\left\langle e^{\beta\lambda\mathbf{d}_\tau^{\rm t}\cdot\mathsf{B}_{\alpha} \cdot\mathbf{z}_\tau+\beta\lambda ku \int_0^\tau{\rm d}t[\mathbf{z}(t)]_x}\right\rangle_{\rm ren}~.
\label{generating}
\end{eqnarray}
Here $\mathbf{q}_\tau^*=\mathbf{z}_\tau+\mathbf{d}_\tau$ and $\mathbf{q}_0^*=\mathbf{z}_0$ are used to get the second line.
$c_\alpha$ is the multiplicative factor independent of integration.  $N_\alpha$ is the normalization factor for the renormalized integral due to the alteration of the initial and final PDF's by $\mathsf{B}_{\alpha}$. The renormalized integral $\langle\cdots\rangle_{\rm ren}$ in Eq.~(\ref{generating}) can be performed by using the cumulant expansion in terms of renormalized correlation functions $\langle\mathbf{z}(t)^{\rm t}\mathbf{z}(t')\rangle_{\rm ren}$~\cite{kwangmoo}; see Sec.~III in SM~\cite{SM}. In the following, we consider the long-time limit, neglecting terms with $e^{-\mathsf{F}\tau}$ and  $e^{-\mathsf{F}^{\rm t}\tau}$.

For the equilibrium case ($u=0$), the generating function is given by $N_\alpha$ only, given in the large $\tau$ as
\begin{equation}
{\cal G}_{\alpha}^{\rm eq}(\lambda)=\sqrt{\frac{|\mathsf{A}_{\rm eq}|^2}{|\mathsf{A}_{\rm eq}+\lambda \mathsf{B}_{\alpha}||\mathsf{A}_{\rm eq}-\lambda \mathsf{B}_{\alpha}|}}=\frac{1}{(1-\lambda^2)^{\nu}}~,
\end{equation}
where $\nu=1,~1+N/2,~N+1$ for $\alpha={\rm S},~{\rm B},~{\rm SB}$, respectively.  Using the Fourier transformation, we evaluate the equilibrium heat distributions for dimensionless heat $r=\beta Q$, given as
\begin{equation}
{\cal P}_{\rm \alpha}^{\rm eq}(r)=
\frac{(|r|/2)^{\nu-1/2}}{\sqrt{\pi}\Gamma(\nu)}K_{\nu-1/2}(|r|)~,
\end{equation}
where $K_{\nu}(z)$ is the second-kind modified Bessel function of order $\nu$. Figure~\ref{fig2} shows a clear difference in three heat distributions depending on $N$, but their averages vanish as expected in equilibrium.
It is interesting to note that ${\cal P}_{\rm S}^{\rm eq}(r)=e^{-|r|/2}$ is independent of $N$ and has been found to be consistent with the equilibrium heat distribution for the single-particle Langevin system~\cite{kwon_langevin,CM}. 

For the nonequilibrium case with $u\neq 0$, we find $\langle W\rangle=ku\tau[ \mathsf{F}^{-1}\cdot\mathbf{u}]_{x}\to N\gamma u^2 \tau$  for large $\tau$.
This is exactly $N$ times larger than the corresponding value for the single-particle Langevin system~\cite{kwon_langevin,CM},
which implies that the dissipation coefficient for the colloid particle increases linearly with the number of interacting bath particles. This is consistent with the usual Stokes' formula~\cite{stokes,donghwan} and indicates that our rather oversimplified model still describes the colloidal particle dynamics reasonably well.

We compute Eq.~(\ref{generating}) for the large $\tau$ limit and find
\begin{equation}
{\cal G}_{\alpha}(\lambda)\simeq\frac{e^{-N\tau w\lambda(1-\lambda)-Nw b_{\alpha}\lambda^3/[2(1+\lambda)]}}{(1-\lambda^2)^\nu}~,
\end{equation}
where $w=\beta\langle W\rangle/(N\tau)=\gamma u^2$ and $b_{\alpha}$'s in unit of time differ for three heats $Q_{\alpha}$; see Sec.~IV in SM~\cite{SM}. 
The heat distribution function for $\beta Q=N\tau w q$ can be obtained by the Fourier integral as
\begin{equation}
{\cal P}_{\alpha}(q)=\int_{-i\infty}^{i\infty}\frac{{\rm d}\lambda}{2\pi i}\frac{e^{-N\tau w[\lambda(1-\lambda)-q\lambda]-Nw b_{\alpha}\lambda^3/[2(1+\lambda)]}}{(1-\lambda^2)^\nu}~,
\label{integral_NEQ}
\end{equation}
which can be evaluated by using the saddle-point approximation due to singularities~\cite{jslee1,jslee2,kwangmoo}. 
The saddle point $\lambda^*$ occurs in the range $-1<\lambda^*<1$. We consider three piecewise regions: (1) far from $\lambda^*=\pm 1$ corresponding to $-1<q<3$ (center); (2) $\lambda^*\simeq 1$ corresponding to $q<-1$ (left wing); (3) $\lambda^*\simeq -1$ corresponding to $q>3$ (right wing). 
After some algebra (see Sec.~V in SM~\cite{SM}), we find
\begin{widetext}
\begin{equation}
{\cal P}_{\alpha}(q)=\left\{
\begin{array}{ll}
\exp\left[-\frac{N\tau}{4} w(1-q)^2-\frac{1}{2}\ln(w\tau)\right]& \textrm{; $-1< q<3$}\\
\exp\left[N\tau w q+(\nu-1)\ln(w\tau) \right]&\textrm{; $q<-1,~|q+1|\gg \left(\frac{Nw\tau}{8\nu}\right)^{-1/2}$}\\
\exp\left[-N\tau w (q-2)+N\sqrt{2w^2b_\alpha\tau(q-3)}+\frac{1}{2}\left(\nu -\frac{3}{2}\right)\ln\left[w^2b_\alpha\tau(q-3)\right]\right]&\textrm{; $q-3\gg \left(\frac{2 \tau}{b_\alpha}\right)^{-2/3}$}
\end{array}\right.~.
\end{equation}
\end{widetext}
The most significant corrections to the large deviation function, proportional to $\tau$ in the exponent of the above  equation, appear in the right wings for $q>3$. Difference arises from the initial memory effect of different Hamiltonians in Eq.~(\ref{3heats})~\cite{farago,puglisi,jslee1,jslee2,kwangmoo}.

For a general strongly coupled system and bath, we find the appearance of the three different heats and various forms of FT involving different heats. From the example of a sliding harmonic potential, we confirm the FT for the total entropy production providing a tighter bound than other FT's and
manifest explicitly the difference in fluctuations of three heats in equilibrium and also nonequilibrium steady states. It would be interesting to study on various heats and FT's for quantum systems.

\begin{acknowledgments}
This work was supported by
the NRF Grants No.~2016R1D1A1A09 (CK), 2017R1D1A1B03030872 (JU), 2017R1D1A09000527 (JY),
and 2017R1D1A1B06035497 (HP), Korea.
\end{acknowledgments}


\begin{references}
\bibitem{exp1}  As there is no membership change of bath particles in interactions, there would be considerable memory remaining, which might cause some unphysical results. However, the qualitative behaviors of the three heat PDF's seem not seriously damaged.
\end{references}
\end{document}